\documentclass[nofootinbib,10pt,twocolumn,pra,showpacs,tightenlines, showkeys, floatfix, english,aps]{revtex4}
\usepackage[T1]{fontenc}
\usepackage[latin9]{inputenc}
\usepackage{color,times}
\usepackage{amsmath}
\usepackage{graphicx}
\usepackage{babel}
\usepackage{esint}
\usepackage{epstopdf}
\usepackage{epsfig}

\makeatletter

\begin{document}

\title{Quantum cryptography over non-Markovian channels}

\author{Kishore Thapliyal$^{a,}$\footnote{Email: tkishore36@yahoo.com}, Anirban Pathak$^{a,}$\footnote{Email: anirban.pathak@gmail.com},
Subhashish Banerjee$^{b,}$\footnote{Email: subhashish@iitj.ac.in}}

\affiliation{$^{a}$Jaypee Institute of Information Technology, A 10, Sector-62,
Noida, UP-201307, India\\
$^{b}$Indian Institute of Technology Jodhpur, Rajasthan-342011, India}

\begin{abstract}
A set of schemes for secure quantum communication are analyzed
under the influence of non-Markovian channels. By comparing
with the corresponding Markovian cases, it is seen that the average
fidelity in all these schemes can be maintained for relatively longer
periods of time. The effects of non-Markovian noise on a number of facets of quantum
cryptography, such as  quantum secure direct communication,
deterministic secure quantum communication and their controlled counterparts,  quantum dialogue, 
quantum key distribution, quantum key agreement, etc., have been extensively investigated. 
Specifically, a scheme for controlled quantum dialogue (CQD) is analyzed over
damping, dephasing and  depolarizing non-Markovian channels, and subsequently, the effect of these non-Markovian channels on the other schemes of secure quantum communication is deduced from the results obtained for CQD. The damped non-Markovian
channel causes, a periodic revival in the fidelity; while fidelity
is observed to be sustained under the influence of the dephasing non-Markovian
channel. The depolarizing channel, as well as the other non-Markovian
channels discussed here, show that the obtained average fidelity subjected
to noisy environment depends on the strength of coupling between the quantum system with its surroundings and the number of rounds
of quantum communication involved in a particular scheme.
\end{abstract}



\maketitle

\section{Introduction}

Quantum cryptography, after its inception in 1984 \cite{bb84}, has
been flourishing over the last decade. The prime reason is the possibility
of unconditional security, a task unachievable in the domain of
classical physics. This fact and already available marketable products
based on quantum cryptography have motivated further research in this
field. To name a few, apart from the initial interest in quantum key
distribution (QKD) \cite{bb84,b92,ekert,vaidman-goldenberg,BBM},
various schemes concerning direct communication (secure communication
circumventing the need of a prior shared key) \cite{ping-pong,lm05,dsqc-1,review,Anindita,cdsqc},
quantum key agreement (QKA) \cite{Qka-CS}, quantum secret sharing
\cite{Hillery}, have been proposed in the recent past (see
\cite{book} for details). Specifically, in the direct communication, the
receiver may or may not require an additional classical information
to decode the message sent by the sender; depending upon this, the
protocol falls under the category of deterministic secure quantum communication
(DSQC) \cite{Anindita} and quantum secure direct communication (QSDC)
\cite{ping-pong,lm05,dsqc-1,review}, respectively. There is another novel
technique of direct communication, quantum dialogue (QD) \cite{ba-an},
where both the users can send their information simultaneously, with
no need of a prior shared key.

All these schemes for secure direct quantum communication, provide us a vast potential
for extension and modification to design protocols required in various
real life scenarios. One such important facet of quantum cryptography
provides solutions for maintaining the hierarchy in offices or government, 
in terms of the information accessible to each user. Hierarchical
quantum communication schemes are aimed to deal with these problems,
when either only single sender holds all the information \cite{HQIS-Shukla}
or it is distributed among two of them \cite{HJRSP}. We may consider another important scenario, 
where a controller supervises the communication among all the remaining users, and he can maintain
his control by making sure that the communication is not accomplished
without his consent \cite{switch,cdsqc,crypt-switch}. Further, a scheme for quantum controlled
communication based on a quantum cryptographic switch has been proposed recently, which
allows the supervisor to control even the amount of information he
wishes to share with the other users in a continuously varying degree
\cite{switch,crypt-switch}.

It would be worth summarizing that the
security achieved in all the cryptographic schemes is based on the
principle of splitting the whole information in many pieces, and the
whole information can only be extracted if all the pieces are available
simultaneously. Usually, one of the parties prepares an entangled
state to be used as a quantum channel and shares it with all other parties
in a secure way. By secure, we mean that a proper eavesdropping checking
technique is employed, after inserting the decoy qubits with the entangled
qubits to ensure the absence of Eve. Once this channel is shared
the legitimate parties can securely share their secrets, either by teleportation
or encoding their information using Pauli operations
and sending the qubits to the receiver again in a secure manner. An
interesting observation, we would like to exploit here, is that if we
start with a controlled quantum dialogue (CQD) scheme, we can
reduce it to almost all the schemes of secure quantum communication. This
point is discussed in detail in the forthcoming sections.

The feasibility of implementation of various quantum communication schemes
when subjected to noisy environment has been analyzed in the past.
In particular, the schemes of QKD \cite{vishal}, QKA
\cite{vishal}, controlled DSQC (CDSQC) \cite{switch}, QSDC \cite{vishal},
CQD \cite{crypt-switch}, QD \cite{vishal}, asymmetric QD (AQD) \cite{AQD},
controlled bidirectional remote state preparation  \cite{vishalrsp}, among others, have been considered under 
the influence of both purely dephasing and damping
noises. Most of these investigations (cf. \cite{vishal,switch,crypt-switch,AQD}) were restricted
to the domain of Markovian environments \cite{pd,sgad}, though, some attempts have been made to
study the effects of non-Markovian environments on quantum communication
schemes, such as teleportation \cite{tel-NM,Tel-NM-IJQI}, densecoding
\cite{tel-NM} and entanglement swapping \cite{EntSwap-NM,NM-MPEntSwap}.
The security of a QKD protocol has also been analyzed over
non-Markovian depolarizing channel \cite{NMDepQKD}. All these attempts
(except Ref. \cite{NMDepQKD}) to examine the usefulness of entangled
states under the influence of non-Markovian environments were restricted
to insecure quantum communication, where security is not required.
However, in the secure quantum communication protocols, it becomes relevant
to differentiate the disturbance caused due to eavesdropping and the
effects of noise. This sets the motivation for this work, where we
wish to analyze the effect of non-Markovian noisy environment considering
the scenario when no eavesdropping has been attempted. This would provide a 
threshold of error due to disturbance from a non-Markovian environment; errors exceeding this could be attributed to the presence of an eavesdropper. 
Specifically, we would consider  pure dephasing, damping and depolarizing interactions with a  non-Markovian 
reservoir. Though entanglement can only be maintained for relatively
longer time due to dephasing non-Markovian interaction \cite{Deph-NM},
it can show revival under dissipative interactions \cite{EntDyn-NM}. As we are essentially using entangled states and
entanglement revival could be an interesting feature to affect the
feasibility of quantum cryptographic schemes, we would like to address
the problem here.

Non-Markovian noise has been attracting a lot of interest from both quantum
optics and quantum information communities, theoretically as well as
experimentally. A paradigm for studying non-Markovian evolution is the quantum Brownian motion
\cite{grabert,sb1,sb2}. Specifically, degradation of purity and nonclassicality
of Gaussian states have been studied under the effect of non-Markovian
channels \cite{NM-Gau}. Dynamics of entanglement has been discussed
in both discrete \cite{NM-qubit,EntDyn-NM,EntDyn-2oscNM,NM-Qjump}
and continuous \cite{NM-CV} variable channels. Recently, dynamics
of multipartite entanglement and its protection have been addressed
\cite{NM-EntProt}. The additional problems due to non-Markovian noise
in quantum error correction \cite{NM-QEC} and dynamical deocupling
\cite{NM-dyDecou-Dy,NM-dyDecoup} have also been discussed in the past.
The non-Markovianity was also characterized from an information theoretic
approach in terms of quantum Fisher information flow \cite{QFI-NM}.
 A number of beautiful experiments depicting non-Markovian nature
of the system-reservoir interaction have been performed \cite{NM-exp,NM-exp-Ph,NMExpScRep}. 

First, the Kraus operators
of non-Markovian dissipative and dephasing noise models  are discussed in a concise manner (in Section \ref {sec:noise-models}).
In Section \ref{sec:effect-of-noise}, we introduce, briefly, a CQD scheme (\ref{sub:CQD})
using Bell states  and based on the quantum cryptographic switch. Then,
we study the effect of non-Markovian noise on the feasibility
of the CQD scheme. To quantify the effect of
noise, a distance-based measure known as fidelity has been calculated. 
Next, we reduce the scheme of CQD to design a CDSQC protocol
(in Section \ref{sub:CDSQC}), a QD protocol (in Section \ref{sub:QD}),
a DSQC and QSDC protocols (in Section \ref{sub:QSDC/DSQC}), a QKA
protocol (in Section \ref{sub:QKA}), and finally, two well known QKD
protocols (in Section \ref{sub:QKD}). The QKD protocols discussed
here are well known as BB84 \cite{bb84} and BBM \cite{BBM} protocols. The feasibility
of all these schemes under the action of non-Markovian channels are also analyzed.
Finally, we conclude the paper in Section \ref{sec:Conclusion}.

\section{Non-Markovian noise models \label{sec:noise-models}}

 We briefly discuss below, a few non-Markovian models that are subsequently used to study the performance of various quantum cryptographic schemes. 
 The dynamics of a system interacting with its surroundings can be expressed
in terms of Kraus operators as 
\begin{equation}
\rho\left(t\right)=\sum_{i}K_{i}\left(t\right)\rho\left(0\right)K_{i}^{\dagger}\left(t\right)\label{eq:noise-affected-density-matrix-1}
\end{equation}
(see \cite{QCh-rev} for a review). Here, we use this approach
to describe the dissipative and purely dephasing interactions with non-Markovian
environments. The Kraus operators for the damping noise under non-Markovian
effects are given by \cite{EntDyn-NM} 

\begin{equation}
K_{0}=|0\rangle\langle0|+\sqrt{p}|1\rangle\langle1|,\,\,\,\,\,\,\,\,\,\,\,K_{1}=\sqrt{1-p}|0\rangle\langle1|,\label{eq:Kraus-damping}
\end{equation}
where $p\equiv p\left(t\right)=\exp\left(-\Gamma t\right)\left\{ \cos\left(\frac{dt}{2}\right)+\frac{\Gamma}{d}\sin\left(\frac{dt}{2}\right)\right\} ^{2}$
with $d=\sqrt{2\gamma\Gamma-\Gamma^{2}}$. Here, $\Gamma$ is the
line width which depends on the reservoir correlation time $\tau_{r}\approx\Gamma^{-1}$;
and $\gamma$ is the coupling strength related to qubit relaxation time
$\tau_{s}\approx\gamma^{-1}$. In the domain of large reservoir correlation
time in comparison to qubit relaxation time, memory effects come into
play. The memory effects are characteristic of non-Markovian nature
of dissipation. Interestingly, taking $p=1-\eta$, the results obtained
for amplitude damping noise under Markovian regime can be deduced,
with $\eta$ being the decoherence rate of amplitude damping channel. 

Similarly, the Kraus operators for purely dephasing non-Markovian noise are \cite{Deph-NM}

\begin{equation}
\begin{array}{c}
K_{0}=|0\rangle\langle0|+p|1\rangle\langle1|,\,\,\,\,\,\,\,\,\,\,\,K_{1}=\sqrt{1-p^{2}}|1\rangle\langle1|,\end{array}\label{eq:Kraus-dephasing}
\end{equation}
where $p\equiv p\left(t\right)=\exp\left[-\frac{\gamma}{2}\left\{ t+\frac{1}{\Gamma}\left(\exp\left(-\Gamma t\right)-1\right)\right\} \right]$.
All the parameters have the same meaning as above. As in the case of dissipative noise, the result
for the well known phase damping channel  can be obtained from Eq. (\ref{eq:Kraus-dephasing})
by considering $p=\sqrt{1-\eta}$. In what
follows, we consider an independent environment for each qubit as
it travels through different channels; a similar assumption has been made in \cite{NM-2qubit,NM-2qubits-2}.

Finally, a non-Markovian depolarizing channel can be described by
the Kraus  operators $K_{i}=\sqrt{\mathcal{P}_{i}}\sigma_{i}$,
where $\sigma_{0}\equiv I$ and $\sigma_{i}$s are the three Pauli
matrices. The $\mathcal{P}_{i}$s should remain positive to ensure
the complete positivity for all values of $\frac{\gamma_{j}}{\Gamma_{j}}$
and are given by \cite{NMDepCh} 
\[
\mathcal{P}_{1}=\frac{1}{4}\left[1+\Omega_{1}-\Omega_{2}-\Omega_{3}\right],
\]
\[
\mathcal{P}_{2}=\frac{1}{4}\left[1-\Omega_{1}+\Omega_{2}-\Omega_{3}\right],
\]
\[
\mathcal{P}_{3}=\frac{1}{4}\left[1-\Omega_{1}-\Omega_{2}+\Omega_{3}\right],
\]
and
\[
\mathcal{P}_{4}=\frac{1}{4}\left[1+\Omega_{1}+\Omega_{2}+\Omega_{3}\right].
\]
Here, $\Omega_{i}=\exp\left(-\frac{\Gamma t}{2}\right)\left[\cos\left(\frac{\Gamma d_{i}t}{2}\right)+\frac{1}{d_{i}}\sin\left(\frac{\Gamma d_{i}t}{2}\right)\right]$
with $d_{i}=\sqrt{16\left(\frac{\gamma_{j}^{2}}{\Gamma_{j}^{2}}+\frac{\gamma_{k}^{2}}{\Gamma_{k}^{2}}\right)-1}$
for $i\neq j\neq k$ \cite{NMDepCh}. Further, $\gamma$ is the coupling
strength of the system and $\Gamma$ is the noise bandwidth parameter.
It should be noted that the Markovian case can be deduced from the
above by taking $\Omega_{i}=\exp\left(-\frac{\gamma_{i}t}{2}\right)$
with $\gamma_{i}=\frac{4}{\Gamma}\left(\gamma_{j}^{2}+\gamma_{k}^{2}\right)$
for $i\neq j\neq k$ \cite{NMDepCh}. 

\section{Effect of non-Markovianity on the Secure Quantum Communication schemes \label{sec:effect-of-noise}}
In what follows, we consider a set of quantum cryptographic protocols and analyze the feasibility of their implementation 
over the  above discussed non-Markovian channels. For all the one-way schemes for quantum cryptography that are discussed here, we consider Alice as the sender and Bob as the receiver, unless stated otherwise; whereas, Charlie is the third party supervising the protocol and referred to as the
controller. However, for two-way schemes (e.g., QD, CQD), both Alice and Bob are considered to play dual roles of receiver and sender.

\subsection{CQD \label{sub:CQD}}

Let us start  with a three party protocol for quantum cryptography, where
two parties (Alice and Bob) wish to communicate simultaneously under
the control of a third party (Charlie). In fact, all the controlled
communication protocols work under an assumption that all the parties
are semi-honest. Otherwise, Alice and Bob can share a quantum state
of their own and circumvent Charlie's control. In literature,  this is sometimes 
viewed as Alice and Bob lacking resources for state preparation, and consequently, they do not set up a quantum
channel between them, rather they rely on Charlie to prepare it for them.

To begin with, we consider a CQD scheme recently proposed by some of the
present authors \cite{crypt-switch}. Charlie prepares $n$ copies of a
Bell state and makes two strings $S_{A}$ and $S_{B}$ of all the
first and second qubits. Subsequently, he sends both the strings to
Bob, only after permuting $S_{B}$\footnote{Here, and in what follows, all the qubits traveling from one party
to other are sent in a secure manner, i.e., to send a sequence of $n$ travel qubits, an equal number of decoy
qubits  are inserted randomly in the original sequence of the travel qubits, and subsequently, these decoy qubits are measured to check the existence of 
eavesdropper(s). Various choices of decoy qubits and the corresponding
principles of security are discussed in \cite{Kishore-decoy}.}. Bob will encode his message by using Pauli operations on the qubits
in string $S_{A}$. Subsequently, Bob sends $S_{A}$ to Alice, who
returns it to him after encoding her secret message as Bob did. It
is pre-decided that Pauli operations $I$, $X,$ $iY$, and $Z$ 
correspond to encoded bit values 00, 01, 10, and 11, respectively.
Finally, Charlie discloses the permutation operator, and using this
information Bob performs a Bell measurement on the partner qubits (Bell
pairs). When Bob announces the measurement outcome, both Alice and Bob
can extract each other's message using their own encoding information and the knowledge of initial Bell state prepared by Charlie. If the choice of Bell state prepared by Charlie is made public, it leads to some leakage, which is often considered to be an inherent characteristic of schemes for QD and its variants. However, such leakage can be circumvented if Charlie chooses the Bell state randomly and sends his choice to Alice and Bob by using a scheme of DSQC or QSDC \cite{AQD}. In fact, the schemes of QD are
the most efficient protocols without involving prior key generation. 

Suppose Charlie started with the initial state $\rho=|\psi\rangle\langle\psi|$,
where $|\psi\rangle\in\left\{ |\psi^{\pm}\rangle,|\phi^{\pm}\rangle\right\} $, and $|\psi^{\pm}\rangle=\frac{|00\rangle\pm|11\rangle}{\sqrt2}$,  $|\phi^{\pm}\rangle=\frac{|01\rangle\pm|10\rangle}{\sqrt2}$.
The transformed density matrix over the noisy channel would become
\begin{equation}
\begin{array}{lcl}
\rho^{\prime} & = & \underset{A_{n},B_{n}}{\sum}\underset{i,j,k,l}{\sum}\left(K_{l}\left(p_{4}\right)\otimes I\right)U_{A_{n}}\left(K_{k}\left(p_{3}\right)\otimes I\right)U_{B_{n}}\\
 & \times & \left(K_{i}\left(p_{1}\right)\otimes K_{j}\left(p_{2}\right)\right)\rho\left(\left(K_{i}\left(p_{1}\right)\otimes K_{j}\left(p_{2}\right)\right)\right)^{\dagger}\\
 & \times & \left(\left(K_{l}\left(p_{4}\right)\otimes I\right)U_{A_{n}}\left(K_{k}\left(p_{3}\right)\otimes I\right)U_{B_{n}}\right)^{\dagger},
\end{array}\label{eq:transformed-rho}
\end{equation}
where $K_{i}$s are the Kraus operators for a specific kind of noise
discussed in the previous section and $U_{j}$s are the Pauli operations
by Alice and Bob with $j\in\left\{ j_{00},j_{01},j_{10},j_{11}\right\} $
for $\left\{ I,X,iY,Z\right\} $. Here, we have used different values of $p_{i}$s corresponding to each operation of the Kraus operator 
(from Eq. (\ref{eq:Kraus-damping}), (\ref{eq:Kraus-dephasing}) or the depolarizing channel) on the initial quantum state as the coupling strength during 
various rounds of the quantum communication is assumed to be different.  It may be noted that the second
summation in the right hand side of Eq. (\ref{eq:transformed-rho})
ensures that the map is positive while the first summation corresponds
to the average over all the possible encoding operations that Alice and
Bob are allowed to perform. Thus, the fidelity that we are discussing here, and
in the rest of the paper, is the average fidelity. Further, here, we have
assumed that the qubits not traveling through a quantum channel are
not affected by noise. There are various distance-based measures
to quantify the effect of noise on the quantum state, such as trace
distance, fidelity, and the Bures distance \cite{with-Adam}. The fidelity of the transformed
density matrix with the quantum state in the ideal situation (i.e., in the absence of noise) would be \cite{QCh-rev} 
\begin{equation}
F=\langle\psi^{\prime}|\rho^{\prime}|\psi^{\prime}\rangle,\label{eq:fidelity}
\end{equation}
where the expected quantum state $|\psi^{\prime}\rangle=U_{A_{n}}U_{B_{n}}|\psi\rangle$.

The fidelity of the quantum state transformed under the damping effect of non-Markovian environment is 
\begin{equation}
F=\frac{1}{4}\left[1+2\sqrt{p_{1}p_{2}p_{3}p_{4}}+p_{1}p_{3}p_{4}\left(2p_{2}-1\right)+p_{3}p_{4}\left(1-p_{2}\right)\right],\label{eq:Damp-psi}
\end{equation}
when the initial quantum state prepared by Charlie is $|\psi^{\pm}\rangle$.
As the choice of initial state is solely a decision of Charlie, an independent
choice of the initial state, i.e., $|\phi^{\pm}\rangle$, would lead to the following expression of fidelity
\begin{equation}
F=\frac{1}{4}\left[1+2\sqrt{p_{1}p_{2}p_{3}p_{4}}+p_{1}p_{3}p_{4}+p_{3}p_{4}\left(p_{2}-1\right)\right].\label{eq:Damp-phi}
\end{equation}

 If the state prepared by Charlie were subjected to a non-Markovian dephasing noise, the fidelity would be 
\begin{equation}
F=\frac{1}{2}\left[1+p_{1}p_{2}p_{3}p_{4}\right].\label{eq:Deph}
\end{equation}
It is interesting to see that the obtained fidelity is independent
of the choice of the initial Bell state by Charlie. This  is also seen in analogous scenarios of
Markovian dephasing noise in 
\cite{AQD,crypt-switch,vishal,HJRSP,Kishore-decoy} and references
therein. If we now consider that the system has evolved under the effect
of a depolarizing channel, then following the above prescription,  the analytical expression 
for fidelity can be obtained as
\begin{equation}
F=\frac{1}{2}\left[1+\Omega_{1}^{4}+\Omega_{2}^{4}+\Omega_{3}^{4}\right].\label{eq:Depolarizing-CQD}
\end{equation}

It is interesting to observe the appearance of fourth order terms in all the non-Markovian fidelities, a
signature of four noisy channels acting on the, four, different rounds of quantum
communication. It should be mentioned here that instead of  sending both
the strings to Bob, Charlie could have sent $S_{A}$ to Alice and
$S_{B}$ to Bob. Subsequently, Alice would have sent $S_{A}$ to Bob
after encoding her message and Bob would have encoded his message before
performing the measurement. The obtained fidelity expressions in this
case turns out to be the same as that of the CDSQC protocol, discussed in the
next subsection. The effect of noise in the case discussed here is
more than that in the case of CDSQC. Making use of this observation,  we analyze the scheme of CQD,
described above,  in detail  as the results obtained in the following subsections can be reduced from it.

Now, we will discuss the fidelities for different scenarios depicted in Eqs. (\ref{eq:Damp-psi})-(\ref{eq:Depolarizing-CQD}), for 
both Markovian and non-Markovian noises. The case of the non-Markovian damping/dephasing channels are also considered
for strong and weak coupling regimes. Specifically, we  obtain results in the strong and weak coupling regimes  over non-Markovian
damping channels $\Gamma=0.01\gamma$ and $\Gamma=0.1\gamma$, whereas
for very high values, such as $\Gamma=5\gamma$, it is found to reduce to Markovian
case.  In the following figures, we have used the notation  NM, M, and $\rm{NM_{S}}$, which correspond to the non-Markovian, Markovian, and non-Markovian 
(under strong coupling strengths) regimes of interactions, respectively. 

\begin{widetext}

\begin{figure}
\includegraphics[angle=-90,scale=1.0]{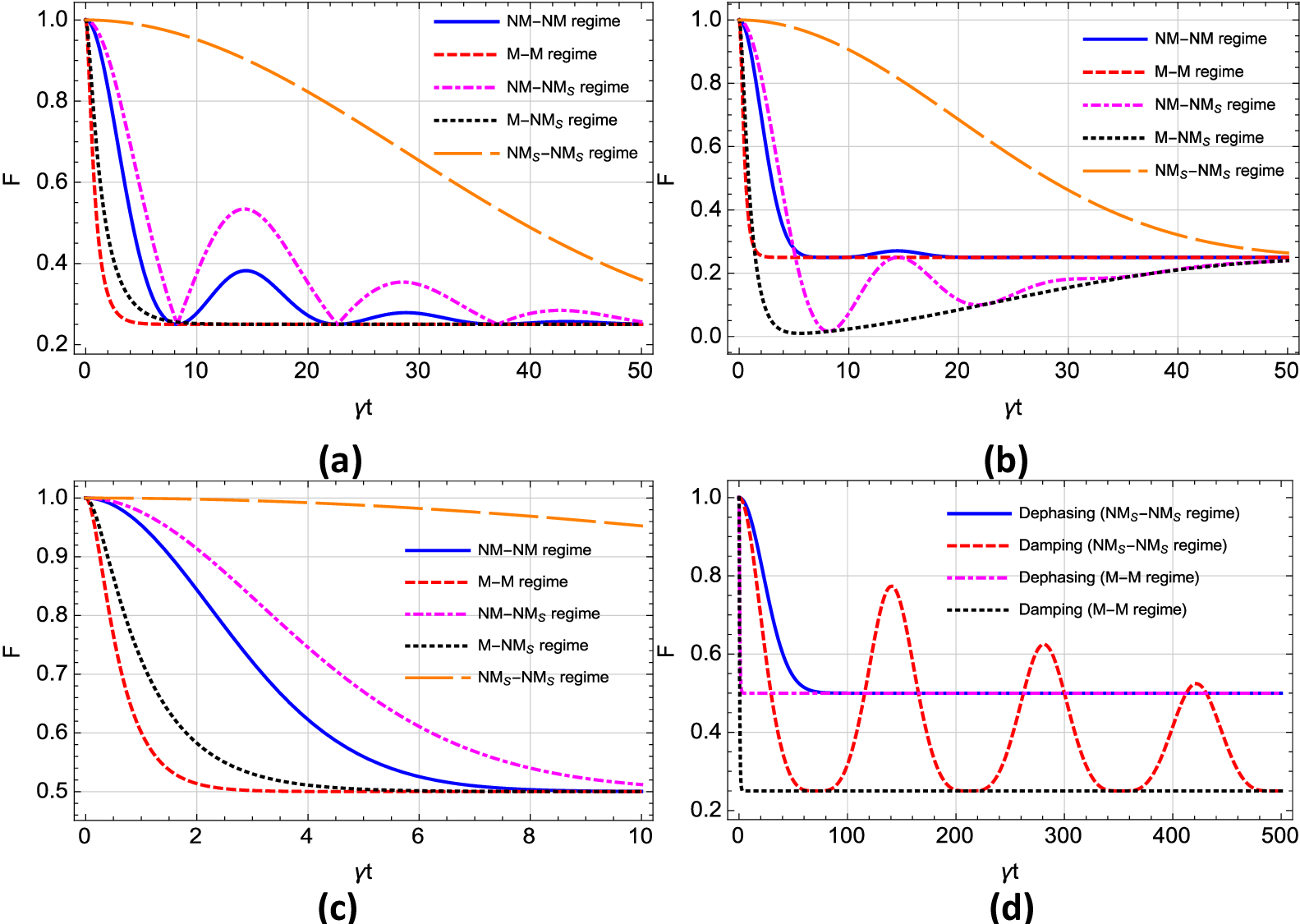}
\protect\caption{\label{fig:CQD-Dam-Deph} (Color online) The variation of the average
fidelity obtained for CQD protocol with respect to the dimensionless quantity $\gamma t$
is depicted when the travel qubits undergo a damping or dephasing interaction
with its surroundings. In (a) and (b), both the travel qubits may have
different coupling strengths during their various rounds of travels
under damping effects, which are characterized by $p_{i}\,:i\in\left\{ 1,2,3,4\right\} $.
The values of coupling strength for strong (weak) regime of non-Markovian
effect is chosen as $\Gamma=0.01\gamma$ ($\Gamma=0.1\gamma$), and
$\Gamma=5\gamma$ for Markovian regime. In (a) and (b), the choice
of initial Bell states by Charlie is $|\psi^{\pm}\rangle$ and $|\phi^{\pm}\rangle$,
respectively. (c) Shows similar cases over the dephasing channels.
In (d), both purely dephasing and damping effects are
shown together for strong non-Markovian and Markovian regimes.}
\end{figure}

\end{widetext}

\begin{figure}
\includegraphics[scale=0.43]{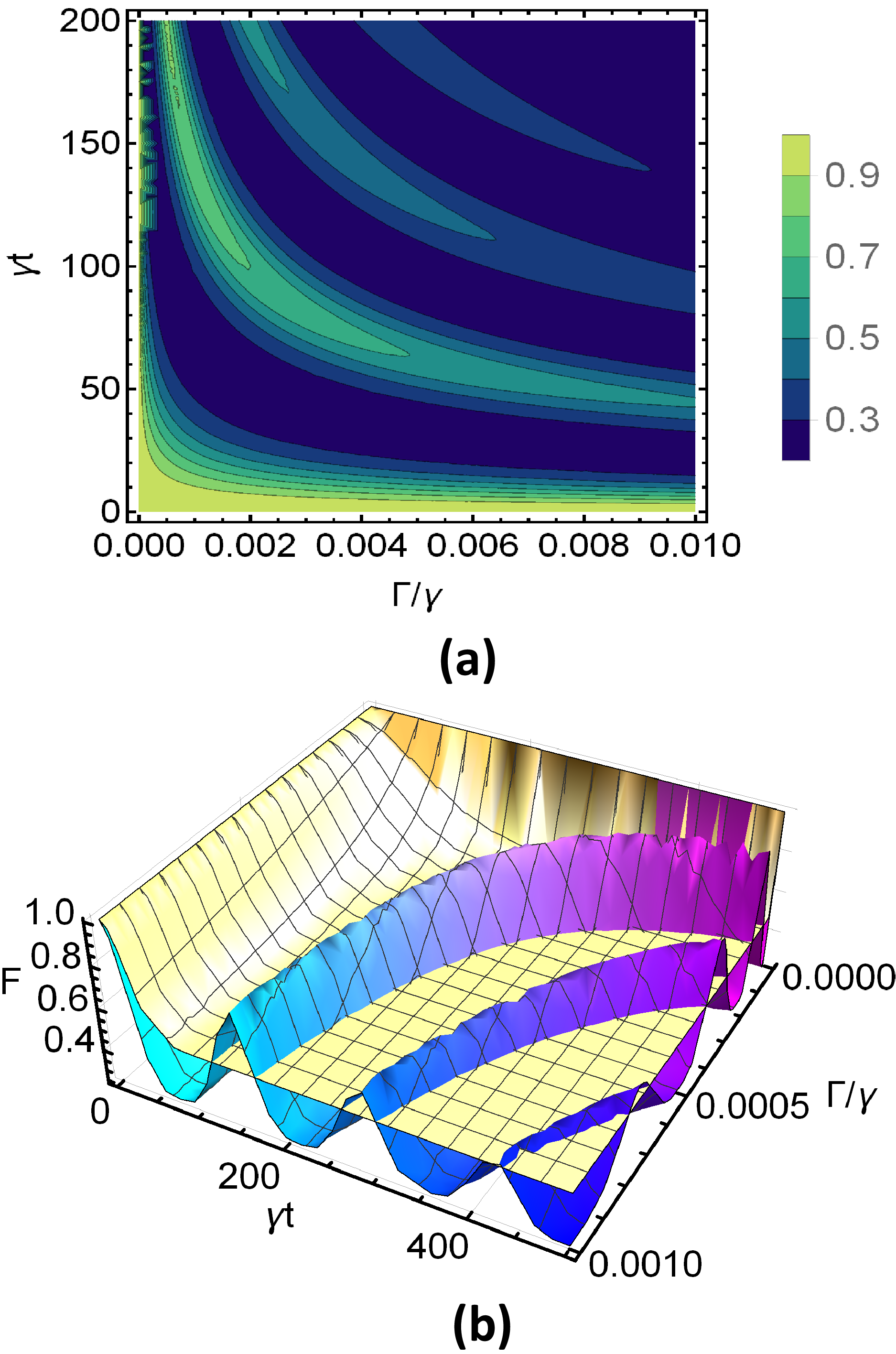}

\protect\caption{\label{fig:CQD-3D}(Color online) The dependence of the obtained fidelity
over the damping channel on the coupling strength and rescaled time is illustrated
through a contour plot in (a). (b) depicts
the variation of the fidelity for varying coupling strength
and time for both purely dephasing and damping non-Markovian channels in light (yellow) and dark (blue) colored surface plots, respectively.}
\end{figure}

\begin{figure}
\includegraphics[scale=0.43]{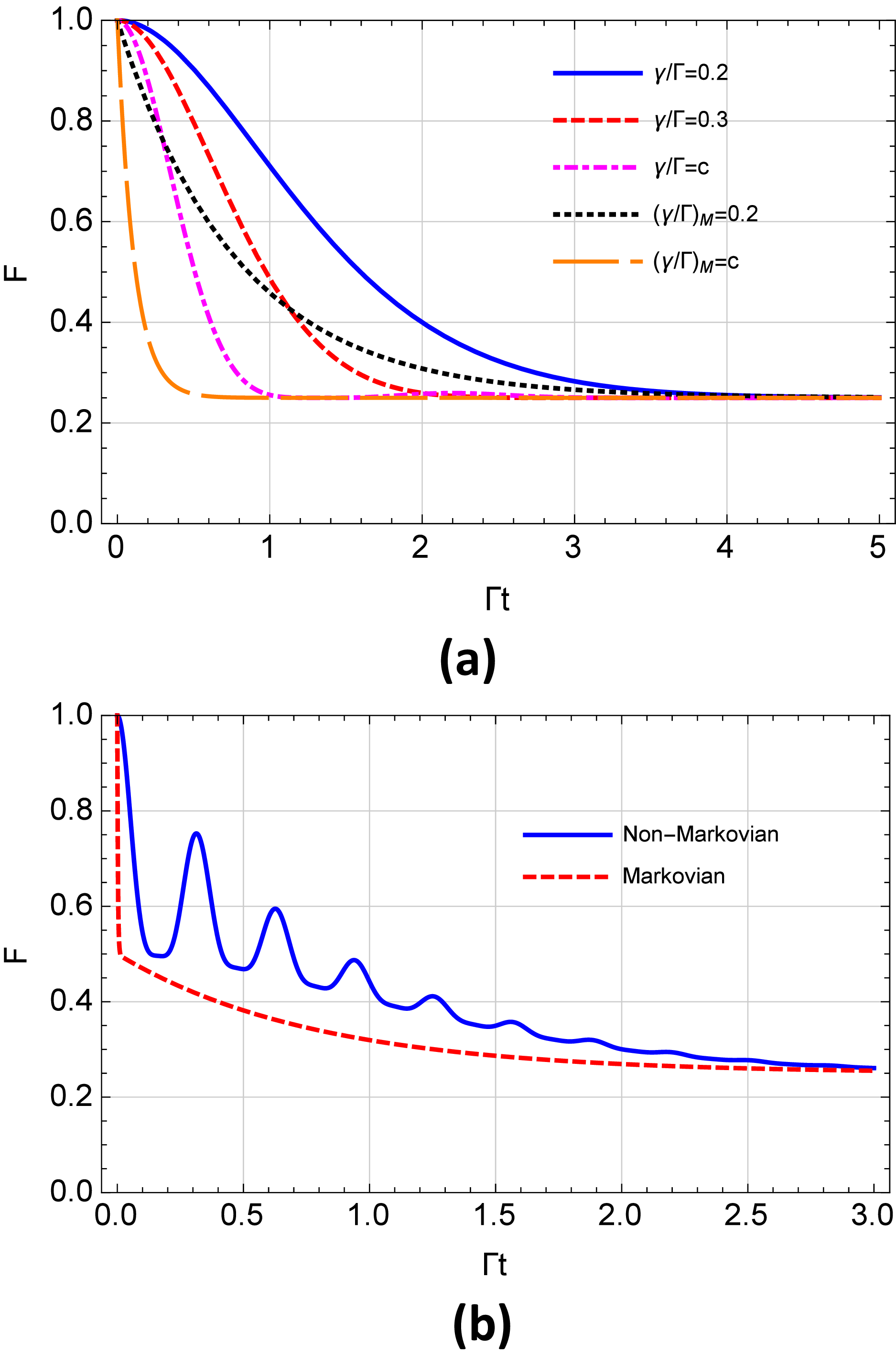}

\protect\caption{\label{fig:CQD-depol}(Color online) (a) The effect of non-Markovian
depolarizing channel on the CQD scheme has been illustrated for different
values of the dimensionless quantity $\frac{\gamma}{\Gamma}$ indicated
in the plot. (a) shows the case of homogeneous non-Markovian depolarizing
channel (i.e., $\frac{\gamma_{i}}{\Gamma_{i}}=\frac{\gamma}{\Gamma}\forall i\in\left\{ 1,2,3\right\} $).
(b) illustrates a comparison between inhomogeneous case of non-Markovian
and Markovian depolarizing channels. In (a), the constant $c=\Gamma\left|\sqrt{\frac{1+\left(\pi/\log3\right)^{2}}{32}}\right|$,
which is the maximum value ensuring completely positive map for all
times for the homogeneous case; in (b), the noise parameters are $\frac{\gamma_{3}}{\Gamma_{3}}=5,\,\frac{\gamma_{i}}{\Gamma_{i}}=0.2$
for $i\in\left\{ 1,2\right\} $.}
\end{figure}

\begin{figure}
\includegraphics[scale=0.43]{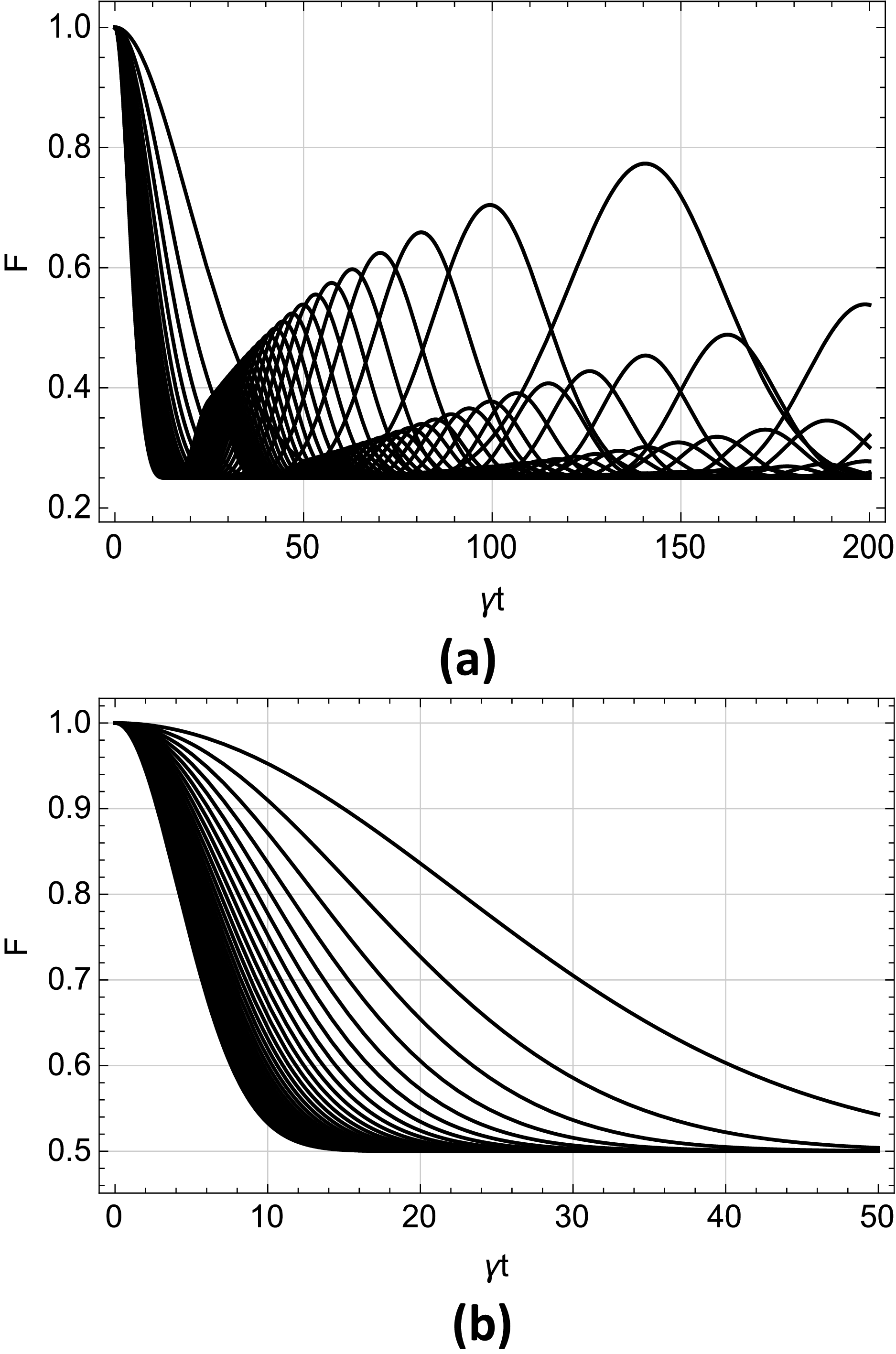}

\protect\caption{\label{fig:CQD-transition-dam-dep} The effect of a change
in the coupling strength on the fidelity is illustrated here
with a set of plots for damping and dephasing non-Markovian noise in
(a) and (b), respectively.  Specifically, the parameter of the coupling strength  $\Gamma/\gamma$ varies from 0.001
to 0.03 in steps of 0.001 in both the plots.  }
\end{figure}

A comparative analysis of the effects of non-Markovian (for both strong and weak
couplings) and Markovian noise can be seen from Figs. \ref{fig:CQD-Dam-Deph}-\ref{fig:CQD-transition-dam-dep}.
In this case, though four different coupling regimes are possible, one for each $p_{i}$s, we have restricted ourselves, for simplicity, to the scenario of 
Charlie to Bob quantum channel having the same coupling strength for both the travel qubits. Similarly, Bob to Alice quantum channel has the same coupling strength as 
that for the other way round. We explicitly mention the two choices of regimes in Fig. \ref{fig:CQD-Dam-Deph}.  Specifically,
Fig. \ref{fig:CQD-Dam-Deph} (a) and (b) show the effect of damping quantified by fidelity on the CQD scheme for different
choices of initial Bell states, i.e., $|\psi^{\pm}\rangle$ and $|\phi^{\pm}\rangle$, respectively. It is interesting to observe that when both the qubits
undergo damping,  either in Markovian or in strong coupling non-Markovian regimes,  the choice of initial Bell states becomes irrelevant (see
red (dashed) and orange (large dashed) curves in Fig. \ref{fig:CQD-Dam-Deph} (a) and (b)). However, this initial choice becomes considerably important
for all the remaining cases, and $|\psi^{\pm}\rangle$ states are seen to be preferable as these states are less affected by 
non-Markovian damping noise than $|\phi^{\pm}\rangle$. Further, it is seen that, due to non-Markovian effects, the fidelity can be maintained for a 
relatively larger period of time (i.e., the quantum state decoheres slowly in non-Markovian environments in comparison to the corresponding Markovian environments), a feature that depends on the coupling strength (cf. Fig. \ref{fig:CQD-Dam-Deph} (a) and (b)). Another interesting characteristic
of this kind of non-Markovian noise is periodicity \cite{EntDyn-NM} and the kinks present in Fig. \ref{fig:CQD-Dam-Deph} (a) and (b) are its signature. 
In Ref. \cite{vishal} it was shown that the dilapidating influence of decoherence, due to Markovian damping, can be checked using squeezing. Here, it is seen that the same 
task can also be achieved by exploiting non-Markovianity. 

The effect of noisy environment is independent of the initial Bell
state over dephasing channel and the fidelity is observed to improve gradually with
non-Markovian effects and coupling strength (cf. Fig. \ref{fig:CQD-Dam-Deph}
(c)). Periodicity in the time variation of fidelity, when all interactions
are (strong) non-Markovian is not visible in the time scale of Fig.
\ref{fig:CQD-Dam-Deph} (a) and (b). For larger time scales, this
can be observed in Fig. \ref{fig:CQD-Dam-Deph} (d), where fidelity
over both the damping and dephasing non-Markovian channels is shown
together. It can be seen that the fidelity under the effect of the
damping noise decays faster than that over dephasing channel. At times,
the fidelity over damping channel is observed to be much larger than that over
dephasing channels, which remains constant at 1/2.

To analyze the effect of the coupling strength with varying time,
we depict, in Fig. \ref{fig:CQD-3D}, a contour and a 3 dimensional plots. The
ripple like plot (cf. the blue-colored surface plot in Fig. \ref{fig:CQD-3D} (b)) shows that with decreasing coupling strength the
amplitudes of the revived fidelity gradually become smaller. The same
fact is also illustrated through a contour plot shown in Fig. \ref{fig:CQD-3D}
(a), where we can see that the area of the light-colored region reduces
as we move from bottom to top. Physically, this corresponds to a transition
from strong to weak coupling non-Markovian regime and finally into
Markovian regime. 

A similar analysis of the fidelity expression for the depolarizing
channel is illustrated in Fig. \ref{fig:CQD-depol}. In Fig. \ref{fig:CQD-depol}
(a), homogeneous depolarizing noise is assumed $\gamma_{i}=\gamma\,\forall i\in\left\{ 1,2,3\right\} $,
for which $\gamma\le\left|\sqrt{\frac{1+\left(\pi/\log3\right)^{2}}{32}}\right|$
to ensure that the dynamical map is completely positive \cite{NMDepCh,NMDepQKD}.
Interestingly, it can be

\noindent seen from Fig. \ref{fig:CQD-depol} (a)
that the fidelity falls gradually with the parameter $\frac{\gamma}{\Gamma}$,
which determines the fluctuation due to the depolarizing channel. However,
it can be noted that for all the cases, the fidelity under non-Markovian
environment is always greater than that for the corresponding Markovian
case, till all the plots merge, with time, to a single value. Further, for the case
of inhomogeneous fluctuations  \cite{NMDepCh,NMDepQKD},
we observe
revival in the fidelity in Fig. \ref{fig:CQD-depol} (b).
From Fig. \ref{fig:CQD-depol}, it can be summarized that non-Markovian
depolarizing channel affects the system less than the corresponding Markovian channel.

The change in coupling strength controls the transition from non-Markovian
to Markovian regime for both damping and depolarizing channels. This
dependence has been illustrated in Fig. \ref{fig:CQD-transition-dam-dep}.
Initial small changes in the value of coupling strength changes considerably the nature
of the obtained fidelity, i.e., the periodicity and
maximum value of fidelity after revival show ample changes for even a
small change of coupling strength. However, for small values of the coupling
strength, this change becomes less sensitive as reflected in the dense black
lines corresponding to smaller values of coupling strengths.

A similar comparison of the effect of non-Markovian
and Markovian depolarizing channels shows that the fidelity sustains
for a longer period of time under the influence of a non-Markovian depolarizing channel, and
is more sensitive to small changes in the noise parameter,
which controls the fluctuation. For higher values
of noise parameter, the variation due to small changes in noise parameters
becomes negligible in both Markovian and non-Markovian depolarizing
channels.

In the following subsections, we will deduce corresponding results for the remaining cryptographic tasks
from the results obtained in this section for the fidelity (for the CQD scheme) over the various non-Markovian channels.

\subsection{CDSQC \label{sub:CDSQC}}

A protocol of CDSQC, based on quantum cryptographic switch,
can be obtained from the CQD scheme discussed in the previous subsection,
i.e., when only a single party encodes and sends his/her message in
a secure manner via the quantum channel, which is decoded by the other
party \cite{cdsqc}. To be precise, Charlie initially follows the
same steps as in Section \ref{sub:CQD} but rather sends the two strings
$S_{A}$ and $S_{B}$ to Alice and Bob, respectively. Alice  encodes
her message as usual and sends the encoded qubit to Bob, who decodes
the secret by performing Bell state measurement on partner pairs with
the help of Charlie \cite{cdsqc}. 

The CDSQC scheme and the effect of noise can be summarized as follows
\begin{equation}
\begin{array}{lcl}
\rho^{\prime} & = & \underset{A_{n}}{\sum}\underset{i,j,l}{\sum}\left(K_{l}\left(p_{4}\right)\otimes I\right)U_{A_{n}}\left(K_{i}\left(p_{1}\right)\otimes K_{j}\left(p_{2}\right)\right)\rho\\
 & \times & \left(\left(K_{l}\left(p_{4}\right)\otimes I\right)U_{A_{n}}\left(K_{i}\left(p_{1}\right)\otimes K_{j}\left(p_{2}\right)\right)\right)^{\dagger},
\end{array}\label{eq:transformed-rho-CDSQC}
\end{equation}
where all the parameters have the same meaning as in Section \ref{sub:CQD}.
It is interesting to observe that the transformed density matrix in
Eq. (\ref{eq:transformed-rho-CDSQC}) can be obtained from Eq. (\ref{eq:transformed-rho})
just by considering $p_{3}=1$ and $U_{B_{n}}=I$. The fidelity can
be calculated with the quantum state expected in the ideal situation,
i.e., $|\psi^{\prime}\rangle=U_{A_{n}}|\psi\rangle$.

Due to this observation, the fidelity of the quantum states affected by the non-Markovian
noise, for the CDSQC scheme can be obtained
from the corresponding CQD expressions by taking $p_{3}=1$,  in
Eqs. (\ref{eq:Damp-psi})-(\ref{eq:Deph}). Interestingly, for the case of the
depolarizing channel, the fidelity can be shown to be 
\begin{equation}
F=\frac{1}{2}\left[1+\Omega_{1}^{3}+\Omega_{2}^{3}+\Omega_{3}^{3}\right],\label{eq:Depolarizing-CDSQC}
\end{equation}
where the presence of cubic terms manifests the fact that the number of rounds of quantum communication involved in this scheme is 
less than that for the scheme discussed in the previous subsection. Specifically, the scheme for CDSQC  requires three rounds of quantum communication, 
while the scheme for CQD requires four rounds.

\begin{widetext}

\begin{figure}
\includegraphics[angle=-90,scale=1.0]{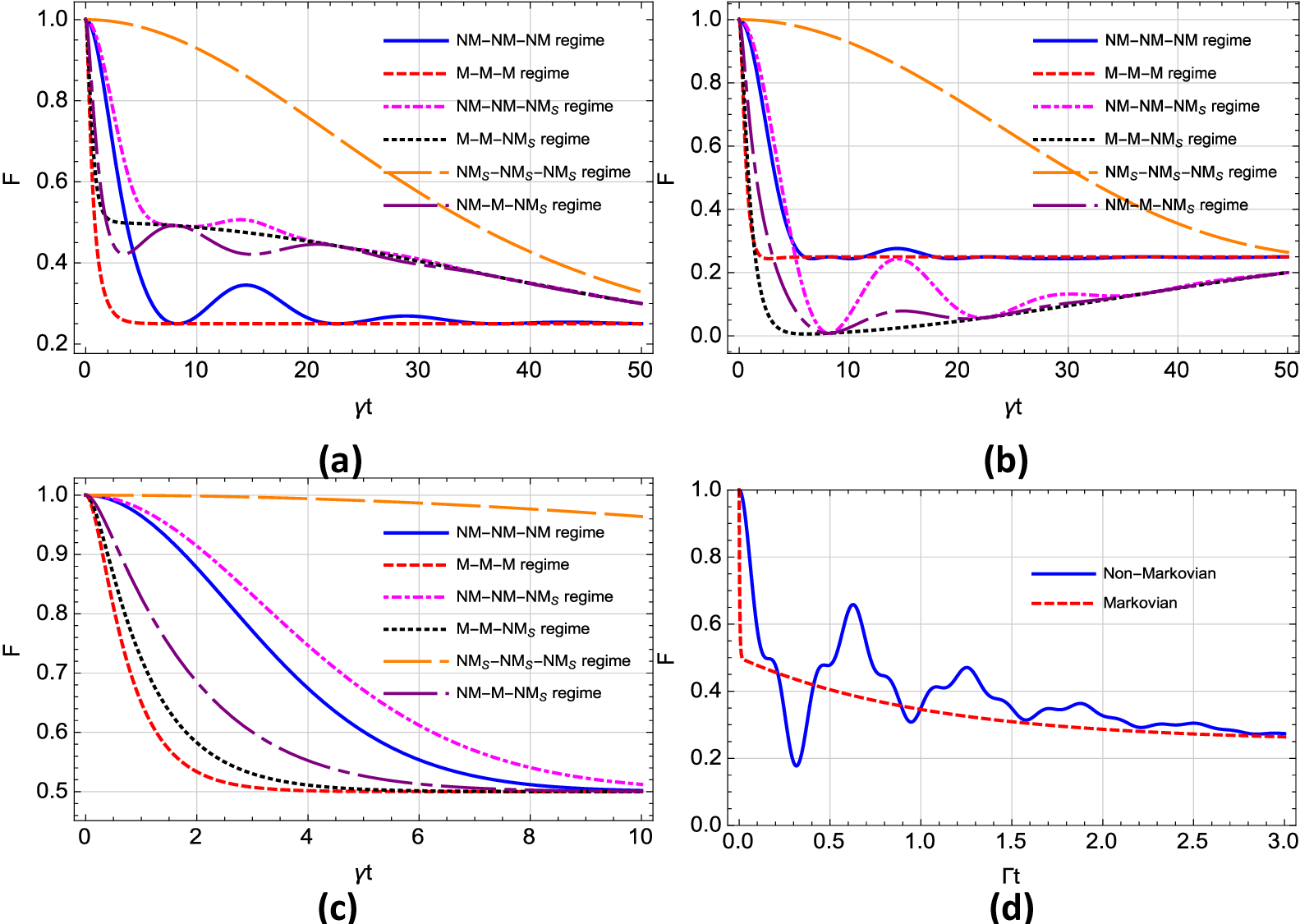}

\protect\caption{\label{fig:CDSQC-Dam-Deph} (Color online) The dependence of the average
fidelity obtained for the CDSQC protocol on the coupling strength is illustrated through its variation with the dimensionless quantity $\gamma t$, when the
travel qubits undergo damping (in (a) and (b)) or dephasing (in
(c)) interaction with their ambient surroundings. In (a) and (b), the initial
state chosen by Charlie was $|\psi^{\pm}\rangle$ and $|\phi^{\pm}\rangle$,
respectively. Here, we have chosen different values of all the coupling
constants in various regimes, i.e., non-Markovian with strong and weak
couplings as well as for the Markovian case. All the values of coupling strengths
corresponding to various regimes are the same as used in the previous
plots. In (d), time variation of the fidelity for the CDSQC
protocol over a depolarizing channel is shown corresponding to the values
used in Fig. \ref{fig:CQD-depol} (b).}
\end{figure}

\end{widetext}

The qubit traveling through the noisy channel may have different coupling
strength during each round of travel. Here, we wish to emphasize this
point with the help of three possible coupling strengths for three
noisy channels acting on the travel qubits. The observations
made above for the extreme cases, i.e., the qubits traveling through either
non-Markovian channels with strong coupling or Markovian channels
all the time, remain valid here as well. Nevertheless, it cannot be conjectured
that the more the number of non-Markovian channels, the higher the fidelity. 
In particular, the large dot-dashed (purple) curve in Fig.
\ref{fig:CDSQC-Dam-Deph} (a) and (b) establishes that  even lower
fidelity is observed with lesser number of Markovian channels acting
on the travel qubits. In fact, Fig. \ref{fig:CDSQC-Dam-Deph} (b)
shows that the obtained fidelity for parity 1 Bell states (i.e., $|\phi^{\pm}\rangle$)
is less for all the cases when various noise channels had different
coupling strengths than that for the case of the travel qubits
subjected to  noisy channels with the same coupling strength.
However, no such nature is visible in Fig. \ref{fig:CDSQC-Dam-Deph}
(c) for dephasing channels.  It is worth stressing here that out of the three possible choices for different coupling regimes corresponding to each $p_{i}$, 
we have emphasized only on the interesting cases and mentioned them accordingly in Fig. \ref{fig:CDSQC-Dam-Deph}. 

Interestingly, the fidelity in the CDSQC protocol falls below the
corresponding Markovian value, under the influence of the non-Markovian depolarizing channel,
when all three noise parameters have different values (cf. Fig. \ref{fig:CDSQC-Dam-Deph}
(d)). This nature can be attributed to the presence of cubic terms
in the fidelity, Eq. (\ref{eq:Depolarizing-CDSQC}).

\subsection{QD \label{sub:QD}}

A CQD scheme can be viewed as a QD scheme under the supervision of a controller.
Therefore, a QD scheme can be easily derived from the CQD scheme if
we consider the scenario that one of the two communicating  parties (i.e., either Alice or Bob) prepares and
measures the quantum state, while both the parties encode their secret on the same
qubits. This QD scheme, which is obtained as a result of reduction from the CQD scheme described above, can be easily recognized to be equivalent to the first QD protocol
proposed by Ba An \cite{ba-an}. The effect of noise on this scheme for QD 
can be obtained  by considering $p_{1}=p_{2}=1$ in all the
expressions of Section \ref{sub:CQD}. This would imply that the initial state is
prepared by one of the communicating parties (say, Bob). Then the transformed density matrix and the fidelity
expressions over non-Markovian channels can be deduced from Eqs. (\ref{eq:transformed-rho})-(\ref{eq:Deph}). Here, it is important to note  that the effect of noise is independent
of the choice of initial Bell state by Charlie/Bob in all the schemes
other than CQD and CDSQC. Similarly, under the effect of depolarizing channels,
the expression of fidelity turns out to be

\begin{equation}
F=\frac{1}{2}\left[1+\Omega_{1}^{2}+\Omega_{2}^{2}+\Omega_{3}^{2}\right],\label{eq:Depolarizing-QD}
\end{equation}
due to two rounds of quantum communication of a travel qubit.

\subsection{QSDC/DSQC \label{sub:QSDC/DSQC}}

As mentioned beforehand in Section \ref{sub:CDSQC}, a CDSQC scheme
can be viewed as a CQD scheme, where only one party is allowed to encode.
In the same way, a QSDC scheme (say, a Ping Pong protocol \cite{ping-pong})
can be viewed as a scheme for QD \cite{ba-an}, where one party (say Bob) is restricted to encode
Identity only. Therefore, all the expressions of the fidelity for a QSDC scheme are exactly the same as those for the scheme of  QD. 

A DSQC scheme can be reduced  from the above mentioned protocols if
Bob incorporates information splitting in two quantum pieces and sends them one after the other 
in two different rounds of Bob to Alice communication \cite{book}. Specifically, Bob prepares two strings
as in Section \ref{sub:QD} and sends the first string to Alice. He
subsequently sends the second string to Alice only if the first quantum
part is received by Alice undisturbed. The effect of non-Markovian
noisy environment on this DSQC scheme can be obtained from the corresponding expressions
for the CQD scheme obtained in Sec. 

\noindent \ref{sub:CQD},
 if we consider $p_{1}=p_{4}=1$
and $p_{2}=p_{3}^{\prime}$ in Eqs. (\ref{eq:Damp-psi})-(\ref{eq:Deph}).
Here, $p_{3}^{\prime}$ is used to show the effect of noise on the
second qubit traveling from Bob to Alice in the first round. In fact,
it turns out to be exactly similar to what is obtained for the QSDC
scheme. Interestingly, in case of depolarizing noise, all the expressions for fidelity 
are found to be the same for QD, QSDC and DSQC schemes. For the convenience
of discussion for the DSQC scheme, we have chosen Bob (Alice) as the sender
(receiver).

So far, we have discussed quantum communication schemes where
prior key generation is circumvented by proper use of quantum resources.
We may now proceed to key generation schemes and investigate the effect of non-Markovian
environment on them.

\subsection{QKA \label{sub:QKA}}

A QKA scheme provides equal power to all the parties taking part
in the key generation process, and does not allow members of a proper subset of the set of all users to solely decide the final
key. Here, we consider a completely orthogonal QKA scheme proposed
in \cite{Qka-CS}. In this QKA protocol, a party (say
Alice) sends her raw key to another party (say Bob) by using a QSDC protocol, while the other party
publicly announces his key. The security of the final key is achieved
by the unconditional security of Alice's transmission of raw key using quantum resources (i.e., from the security of the QSDC/DSQC scheme used by Alice and Bob for Alice to Bob communication). Specifically, Alice transmits a key $k_{A}$ to Bob in a secure manner, whereas Bob announces his key $k_{B}$, publicly, and for all future communication they use a key $k_{AB}=k_{A}\oplus k_{B}$, where $\oplus$ denotes a bitwise XOR operation. Although, Eve knows $k_{B}$, she cannot obtain any information about $k_{AB}$ as she knows nothing about $k_{A}$. Thus, the security of $k_{AB}$ depends on the security of $k_{A}$. In other words, unconditional security of the QSDC scheme involved here would ensure the security of the protocol for QKA.
Interestingly, in Ref. \cite{vishal}, the present authors had already
shown that the effect of noise on this scheme is identical to the
QSDC scheme discussed in the previous Section \ref{sub:QSDC/DSQC}. Since the observations made there remain valid here, we do not  discuss it in further detail.

\subsection{QKD \label{sub:QKD}}

Any discussion on quantum cryptography remains
incomplete without discussing a protocol that changed the course of
cryptography by establishing the feasibility of unconditional security.
In this section, we  discuss two QKD protocols, which can be viewed as the
variants of the same scheme, differing only in the measurement
procedure. Specifically, the BB84 \cite{bb84}
and BBM \cite{BBM} QKD protocols are discussed here. Before we proceed further, it would be apt to note that in contrast to the fidelity expressions obtained in the earlier sections (which were average over all the encoding operations), for QKD protocols, the average fidelity is obtained over all possible equally probable measurement outcomes.

In the BBM protocol \cite{BBM}, Alice prepares $n$ Bell states and sends
all the first qubits to Bob, and both of them measure the qubits of the shared
Bell states randomly in computational ($\left\{ |0\rangle,|1\rangle\right\} $)
and diagonal ($\left\{ |+\rangle,|-\rangle\right\} $) basis. Using
the outcome of these measurements, they finally obtain an unconditionally secure quantum key for those cases where
both Alice and Bob perform measurement using the same basis.

The BB84 protocol can also be viewed along the same lines, where Alice
first measures her qubit (i.e., second qubit) of each Bell state randomly either in computational
or diagonal basis and then sends the other qubit to Bob. Finally,
they can obtain a key by using the measurement outcomes of half of those cases, where they have chosen the same basis. The other half of the cases should be used 
for eavesdropping check. Specifically, when Alice and Bob have performed measurement in the same basis, in the absence of Eve, 
their measurement outcomes should match and a mismatch would indicate the presence of Eve.

Interestingly, for the BBM protocol, the effect of noise can be considered
by taking $p_{1}=p_{2}=p_{4}=1$ in Eqs. (\ref{eq:Damp-psi})-(\ref{eq:Deph}).
Similarly, the effect of depolarizing channel reduces the fidelity
to

\begin{equation}
F=\frac{1}{2}\left[1+\Omega_{1}+\Omega_{2}+\Omega_{3}\right].\label{eq:Depolarizing-QKD-BBM}
\end{equation}
A similar study for the BB84 protocol results in the following fidelity
over damping non-Markovian channels

\begin{equation}
F=\frac{1}{4}\left[2+\sqrt{p_{3}}+p_{3}\right],\label{eq:Damp-QKD}
\end{equation}
while, for the dephasing channel the fidelity is 

\begin{equation}
F=\frac{1}{4}\left[3+p_{3}\right].\label{eq:Deph-QKD}
\end{equation}
Further, the fidelity when the travel qubit is
subjected to a depolarizing channel is 

\begin{equation}
F=\frac{1}{2}\left[2+\Omega_{1}+\Omega_{3}\right].\label{eq:Depolarizing-QKD-BB84}
\end{equation}

Additionally, the present results can also be used to deduce the fidelity for a few other quantum cryptographic schemes, which will
reflect quantitatively the effect of non-Markovian channels on the
corresponding scheme. For example, the effect of noise on Ekert's
QKD protocol \cite{ekert} can also be deduced from the results in
Sec. \ref{sub:CQD}, by taking $p_{3}=p_{4}=1$ as the source
of entanglement is between both the parties, and both the entangled
qubits travel to Alice and Bob from there. Similarly, the feasibility
of the B92 protocol \cite{b92} can also be analyzed over the non-Markovian
channels in analogy with the study for BB84 protocol by only considering
two of the four single qubit states (one each chosen from computational and diagonal basis).

\begin{figure}
\includegraphics[scale=0.43]{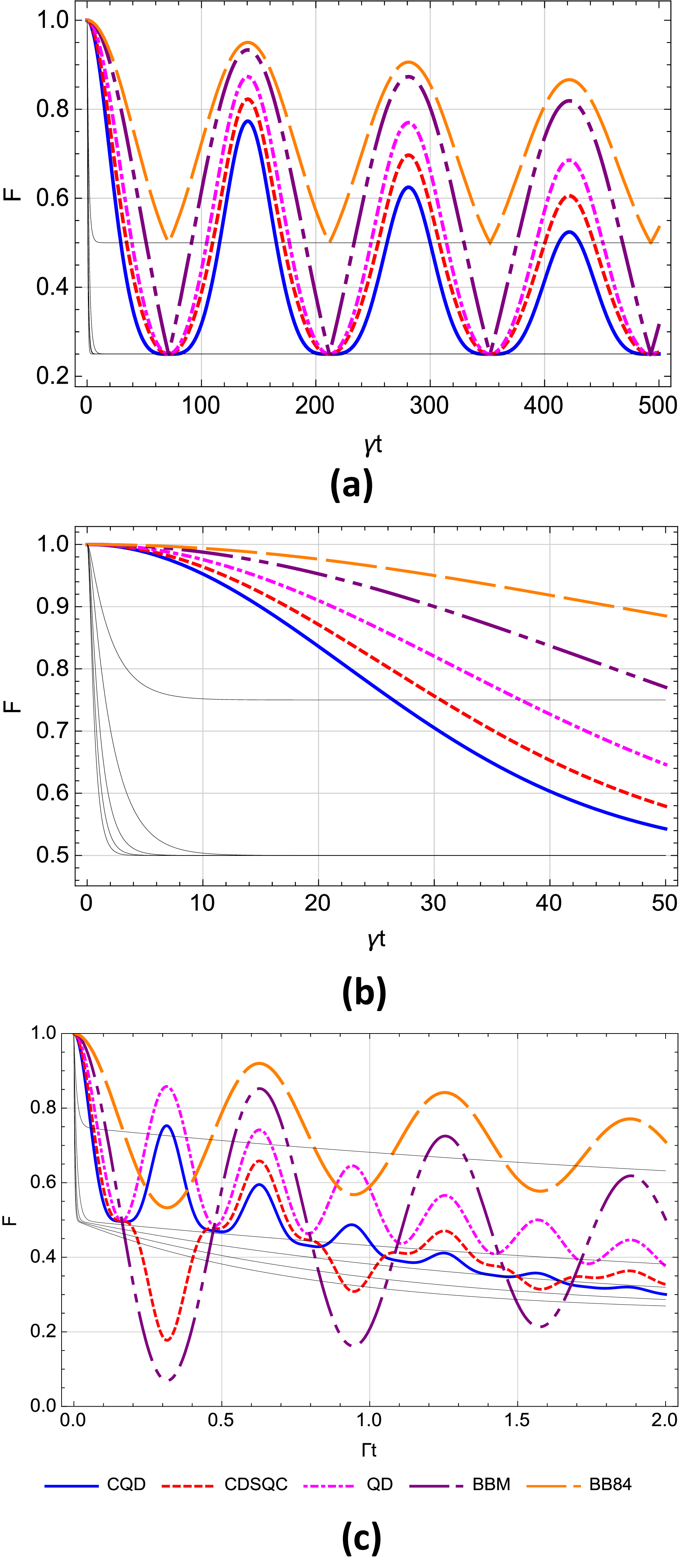}

\protect\caption{\label{fig:All-schemes} (Color online) A comparative analysis of
all the quantum cryptographic schemes discussed so far over the non-Markovian
channels. Each line in all three plots corresponds to the different
cryptographic scheme mentioned in the plot legend at the bottom of
the figure. The light black lines in all three plots represent the corresponding
Markovian cases, and the black lines from bottom to top show the average
fidelity for CQD, CDSQC, QD, BBM QKD, and BB84 QKD protocols. The
fidelity obtained for QSDC, DSQC, and QKA schemes is exactly the same
as that of the QD protocol. }

\end{figure}

Finally, we  perform a comparative study for the fidelity
obtained in each cryptographic scheme to reveal the general
nature of the effect of non-Markovian channels on all these schemes
(shown in Fig. \ref{fig:All-schemes}). Interestingly, the effect
of noise depends on the number of rounds a qubit is required to travel through the noisy
channel. This fact is consistent with the recent observations
on a set of Markovian channels \cite{vishal}. Specifically, in the
CQD scheme, one qubit travels from Charlie to Bob, while another qubit
travels from Charlie-Bob-Alice-Bob. Therefore, the maximum number of rounds
of travels in the set of secure quantum communication schemes discussed
here is four for CQD scheme, which decreases to three for CDSQC. It
further reduces to two for QD, QSDC, DSQC, and QKA schemes. The same
fidelity for all these schemes further establish
this point. Finally,  BBM and BB84 QKD protocols require only
one round of quantum communication. In fact, BBM and BB84 protocols use entangled and single qubit states, respectively, to accomplish
the same task. Out of these two schemes, the BB84 QKD scheme is least affected by noise
as it uses single qubit states, which were shown to be less affected
due to Markovian channels in \cite{vishal}.

In Fig. \ref{fig:All-schemes} (a) and (b), the fidelity
variation over non-Markovian channels due to the strong coupling of
the travel qubits with the environment is depicted. Similarly, the
effect of different noise parameters corresponding to depolarizing channel
is shown in Fig. \ref{fig:All-schemes} (c). Also shown is the
effect of Markovian environment on the fidelity in all three
cases, depicted by  thin smooth (black) lines. For all cases of
Markovian dynamics, the observation that the effect of noise depends
on the rounds of quantum communication remains valid. 

From Fig. \ref{fig:All-schemes} (a), the revival in the fidelity
over non-Markovian damping channel is seen to decrease with an increase in the
number of travel qubits. Similarly, the fidelity falls with
increasing rounds of quantum communication when subjected to dephasing
non-Markovian channel, as shown in Fig. \ref{fig:All-schemes} (b).
Out of the set of fidelities, over the depolarizing channel, those having
odd power terms, such as for the CDSQC and QKD protocols, show
fidelity less than that for the corresponding Markovian case. Otherwise,
in all the remaining cases, the fidelity over non-Markovian
channels is more than that for the corresponding Markovian channels (cf. Fig. \ref{fig:All-schemes} (c)).

\section{Conclusion \label{sec:Conclusion}}

The present study on the effect of a set of non-Markovian channels
on various schemes for secure quantum communication tasks led to a number of interesting
results. Specifically, we have considered here a damping, a purely
dephasing and a depolarizing non-Markovian channel to analyze the feasibility
of some quantum cryptographic schemes evolving under the influence of the non-Markovian environments.
We have started with a CQD scheme, based on a quantum cryptographic switch that uses Bell
states. Later, this scheme is modified to deduce the results for other
quantum cryptographic tasks, such as, CDSQC, QSDC, DSQC. Apart from
these direct communication schemes, the effect of non-Markovian noise on some protocols of QKD and QKA is also
analyzed. 

It has been established that the effect of non-Markovian noise 
depends on the number of rounds of the travel qubits. We have observed that the BB84 QKD scheme is least affected due to
non-Markovian channels, while the CQD scheme shows a maximum fall in
the fidelity. In fact, from the results obtained here one
can also show that the AQD scheme \cite{AQD} will have the same effect
as that on the QD protocol if the number of travel qubits is kept unchanged.
This fact is consistent with the results obtained here, that the fidelity for QSDC, DSQC, and QKA
schemes are exactly the same as that for the QD protocol. In the recent past, we have established that squeezing is a useful quantum resource for 
quantum cryptography as it can help to stop decoherence. Here, we have shown that non-Markovianity can also be used to accomplish a similar task. 

Interestingly, the effect of noise on the CQD and CDSQC schemes is found to depend
on Chalie's initial choice of the Bell state, while it is independent
of this in all the remaining schemes. Finally, our analysis has also
revealed that the fidelity obtained in the case of damping and dephasing
channels depends on the coupling strength. 
 We hope these results would bring out the importance and utility of the non-Markovian behavior in the understanding of quantum cryptographic protocols from the perspective of their practical implementation.

\textbf{Acknowledgment: }AP and KT thank Defense Research \& Development
Organization (DRDO), India for the support provided through the project
number ERIP/ER/1403163/M/01/1603. SB acknowledges support
by the project number 03(1369)/16/EMR-II funded by Council of Scientific
and Industrial Research, New Delhi.

\end{document}